# Electronic Structure and Optical Properties of Monolayer $ReS_2$ with Defect Controlled by Strain Engineering

Yimeng Min, and Lizhe Liu[§],*

[†] *Kuang Yaming Honors School, Nanjing University, Nanjing 210093, P. R. China*

[§]*Institute of Acoustics and Collaborative Innovation Center of Advanced Microstructures, National Laboratory of Solid State Microstructures, Nanjing University, Nanjing 210093, P. R. China*

**Abstract**

By using first-principles calculations, we investigated the monolayer $ReS_2$ with vacancies under strain engineering, specifically focusing on its energy of formation, band gap, electron density of states, effective mass and optical properties. The calculated results disclose that S4 defect is more likely to form than other kinds of vacancies. Asymmetric deformation induced by strain makes its band structure transformation from direct band gap to indirect band gap. The analysis of the partial density of states indicates that the Re-d, Re-p and S-d orbitals are the major components of the defect states, being different from $MoS_2$, the defect states locate both above and below the Fermi level. Moreover, the effective mass was sensitive and anisotropic under the external strain. The reflection spectrum can be greatly tuned by the external strains, which indicates that the $ReS_2$ monolayer has promising applications in nanoscale strain sensor and conductance-switch FETs.

# INTRODUCTION

Many people consider two dimensional (2D) materials like graphene[1–3] and transition metal dichalcogenides (TMDs)[4–7] as candidate materials for future nanoelectronic application because their unusual properties (physical, optical, and electronic) in recent years. By exhibiting a sizable band gap, some 2D TMDs make them be capable to incorporate themselves into digital circuits.[8,9] The investigation on these 2D TMDs like $MoS_2$ and $WS_2$ monolayers has caught many people's eyes. To obtain anisotropic physical characteristics, a lot of attention is attracted by a new 2D TMD: $ReS_2$.[10–15] Photoluminescence measurements showed that the monolayer $ReS_2$ is a direct gap semiconductor (1.55 eV). Also, monolayer $ReS_2$ does not exhibit a direct-indirect gap transition when going from monolayer to bulk, which suggests that this material may be able to become a better candidate on optoelectronic devices comparing to other 2D materials ($MoS_2$)[10]. Using first principle calculations, the effects of doping and defects on the electronic and magnetic properties of $ReS_2$ monolayer have been investigated.[16,17] In addition, the idea that electronic structures and the magnetic properties can be tuned by surface adsorption has been supported by many previous reports on graphene and $MoS_2$.[18–25] For example, people are able to increase the free carrier concentration in $MoS_2$ by absorption and the local magnetic moment can be induced using the adsorption of nonmetal element.[23–25] All those results indicate that defects and deformation play a critical role in their novel characteristics. However, the defect structures in $ReS_2$ monolayer are not clarified, which hinder the application of many field.

In this work, we systematically investigated monolayer $ReS_2$ with sulfur defect under different strains along two directions. We explicitly explored the energy of formation of different kinds of defects of the monolayer by first-principle calculations. Our results showed that one type of defect structure is the most stable, though different deformation is exerted by external strain. We then investigated the electronic properties of this type of defect, the results indicate that deformations induced by strain not only regulate its band structure from direct to indirect, but also lead to flexible tenability of band gap. Partial densities of states as well as the effective mass are calculated, which reveal that the strain applied in different directions can also induce different optical properties (reflection and absorption).

## METHODS

The theoretical assessment is based on the density functional theory in Perdew-Burke-Ernzerh of (PBE) generalized approximation (GGA), using the CASTEP package code with projector augmented wave pseudopotenials. The plane-wave energy cutoff of 400.0 eV is used to expand the Kohn-Sham wave functions and relaxation is carried out until convergence tolerances of $1.0 \times 10^{-5}$ eV for energy. A vacuum space between neighboring layers is set to be more than 25 Å to avoid the interactions between layers. The Monkhorst-Pack k-point meshes (in two-dimensional Brillion zone) of $3\times3\times1$ for the monolayer $ReS_2$, which has been tested to converge.

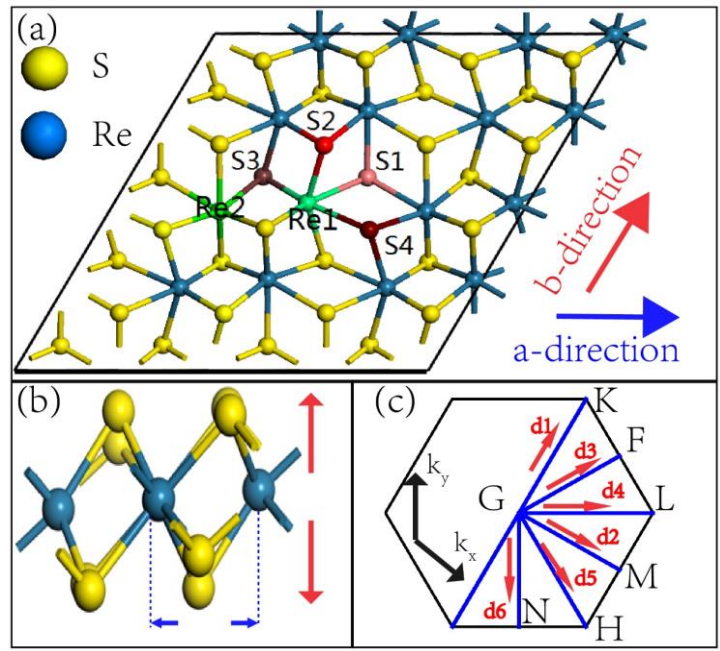

**Figure 1.** (Color on-line) (a) Four different types of Sulfur vacancy and two types of Rhenium in the monolayer $ReS_2$, labeled as S1, S2, S3, S4 and Re1 and Re2. (b) The side view of monolayer $ReS_2$. (c) The first Brillouin Zone (BZ) of monolayer $ReS_2$.

## RESULTS AND DISCUSSION

According to the surface symmetry, there are four types of Sulfur vacancies[26] and two types of Rhenium vacancies in the monolayer ReS$_2$. A typical schematic is displayed in Fig. 1(a), in which 6 different kinds are characterized. The side view of monolayer ReS$_2$ is presented in Fig. 1(b). The special feature is that, unlike hexagonal TMDs (transition-metal dichalcogenides), the S atoms are not all in the same plane, substantially lowering the structure symmetry. The first Brillouin zone (BZ) is plotted in Fig. 1(c), which is a hexagon but with unequal side lengths as a result of the distorted atomic structure. Consequently, the K, L, and H points are no longer equivalent. The vacancy formation energy (EOV) can be expressed as[33]:

For Re: EOV = E$_{vac}$-E$_{st}$ + E$_{Re}$ ; For Sulfur: EOV = E$_{vac}$-E$_{st}$ + 1/32 E$_{S32}$

where E$_{Re}$ is the energy of an isolated Re and E$_{S32}$ is the energy of S$_{32}$ molecule. (which are -303.46 eV and -8880.28eV respectively in our calculation) E$_{st}$ and E$_{vac}$ stand for the energy of the stoichiometric structure and that of the same structure except containing an Sulfur or Rhenium vacancy. Then we define the applied strain rate in a and b directions: $\alpha_a$ and $\alpha_b$

$$\alpha_a = \frac{x - x_0}{x_0}$$

$$\alpha_b = \frac{y - y_0}{y_0}$$

Where x and y are the lengths of the supercell in the a and b directions after deformation, x0= 13.13562Å and y0= 12.799754 Å are the lengths of the supercell in the a and b directions before deformation.

When the external strains are applied along a and b directions [as shown in Fig. 1(a)]. The EOV of Re and S show different trends. For Re vacancies, we can see that the EOV increases slowly with compressed deformation from $\alpha_a = -6\%$ and reaches a maximum value at $\alpha_a = 0\%$, then decreases, the EOV is symmetrical with respect to $\alpha_a = 0\%$. This behavior is different from the sulfur vacancies of Fig. 2, as shown that the deformation of sulfur vacancies along a or b directions leads to an increase of EOV. Those obvious behavioral difference induced by external strain is strongly relevant to their lattice structure. When the strain is applied along a-direction, for Re vacancies, the different strains make the electronic wave functions overlap and develop into more stable status regardless of sketch or squeeze, as a result, the EOVs decrease when external strain is applied; for S vacancies (SV), the smaller the lattice constant along a-direction is, the closer the electronic wave function will overlap and more stable statues will develop into. The condition is similar when the strain is along b-direction, except the S1 vacancy and Re2 vacancy at -4%, these small fluctuations suggest additional formation of bonds. For type S2 and S3, the EOVs of them are close to each other. For type S4, the EOV is always the lowest among all the different vacancies, which suggests that this type of SV is more likely to form in the experiment.

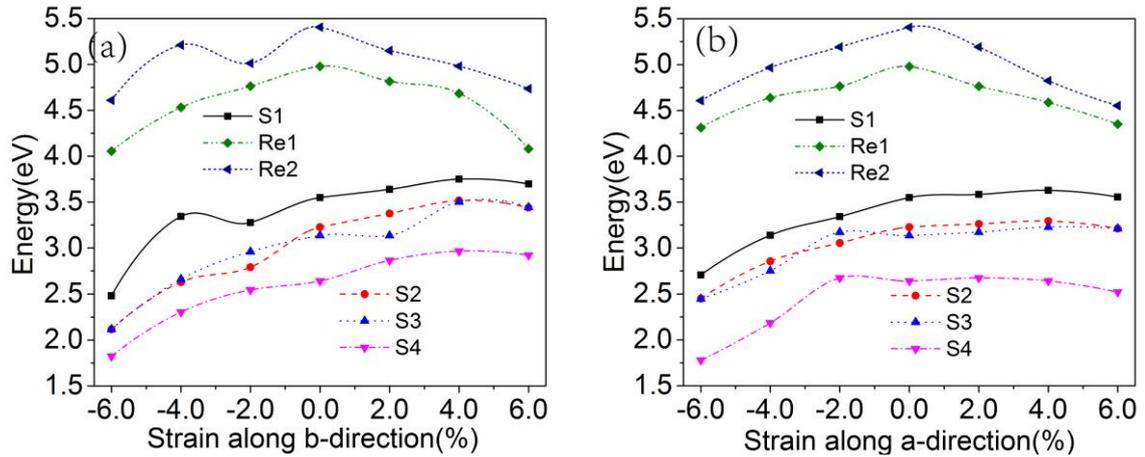

**Figure 2.** (Color on-line) (a) EVO varies with the strains in a-direction. (b) EVO varies with the strains in b-direction.

It is known that different vacancies could convert into each other[27] for lowering the system's energy, which can lead to a finite or even short lifetime of the vacancies in 2D ReS$_2$, depending on their diffusion coefficient. So it is interesting to calculate the diffusion barriers of the stable ground-state S4 in the ReS$_2$. First, since the S4 has the lowest ground state energy, it suggests that S4 is most stable among the SVs and other kinds of SVs may be able to migrate to it. Several possible migration paths have been tested, from which the most favorable path is found, corresponding to the migration process with the minimum number of broken Re-S bonds and S-S bonds. Fig. 3(d) shows the different migration processes. The transition energy barrier (TEB) from S3 to S4 is much larger than the TEB of S1 to S4 and S2 to S4. This can be interpreted by the fact that additional bonds were formed during the migration (covalent bond). The possible transition states for S1 to S4, S2 toS4 and S3 to S4 are shown in Fig. 3(a), Fig.3(b) and Fig. 3(c) respectively, for the transition states of S1 to S4 and S2 to S4, the shortest distance for between to sulfur atoms around the vacancy is about 3.18 Å, however, the distance becomes 2.25 Å in Fig. 3(c). This suggests an additional S-S covalent bond is found to form while migrating from S3 to S4, which indicates that the Sulfur atom needs more energy to get rid of the covalent bond and a higher TEB.

Next, its diffusion coefficient is roughly obtained by using the formula of $D \sim ga^2v_0 \exp(-\frac{E}{k_BT})$, where D is the diffusion constant, $g \sim 1$ is a geometrical factor, a is the single hop distance (here, a = 1.71 Å, 2.92 Å and 3.86 Å respectively.), E is the calculated energy barrier, $k_B$ is the Boltzmann constant, and T is the temperature. And $v_0 = (\frac{k_BT}{2\pi\mu b^2})^{\frac{1}{2}} \sim 5.2*10^{11} s^{-1}$ is the attempt frequency, where $\mu = m_s$ is the mass of a sulfur atoms and b is the equilibrium Re-S bond length (b=2.16 Å). The energy barriers are 2.16 eV, 1.84eV and 7.09eV for S1 to S4, S2 to S4 and S3 to S4 respectively. To make the migration happen, the diffusion coefficient for usual condition[27] is about 10$^{-5}$ cm$^2$/s, for ReS$_2$, it needs to be warmed to over 800 K to reach that standard. Since the diffusion coefficient in the 2D ReS$_2$ is much smaller than other conditions at certain temperature range, the sulfur vacancies will not migrate to the transition state and leads to a long or even infinite lifetime of the SVs in monolayer at a certain temperature range[27].

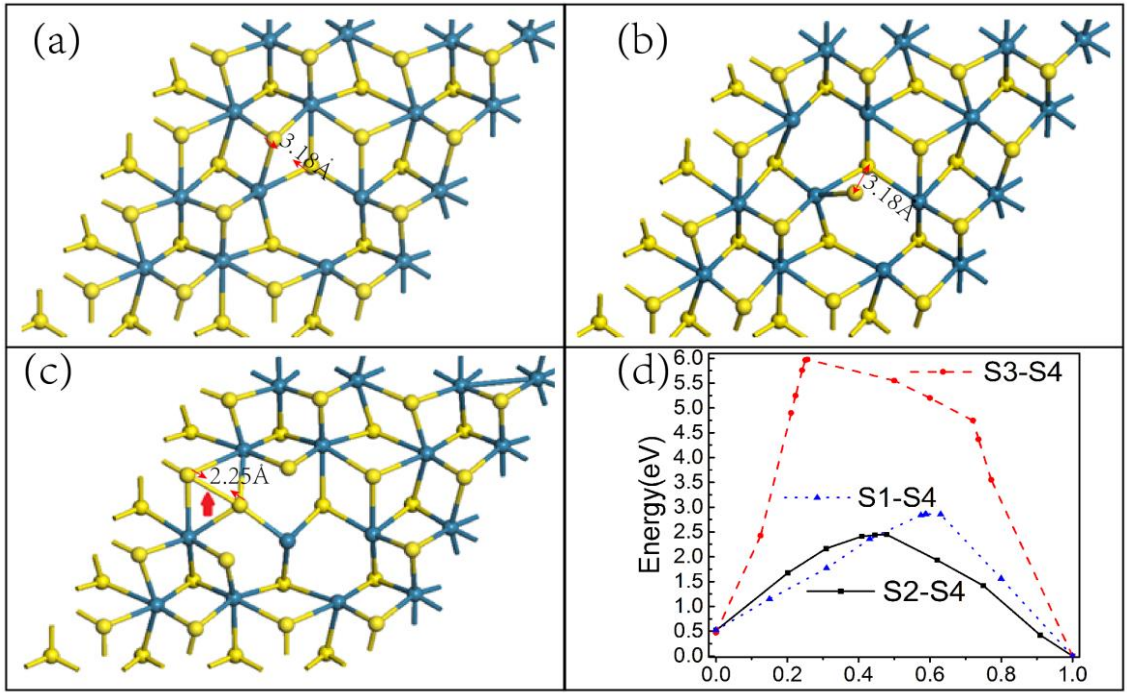

**Figure 3.** (Color on-line) (a) The possible transition state when S1 migrates to S4. (b) The possible transition state when S2 migrates to S4. (c) The possible transition state when S3 migrates to S4. (d) The corresponding transition energy barriers of different SVs.

In order to verify the stability of the SVs at finite temperatures, we have performed the ab initio MD simulations at several different temperatures. It is found that no rotation and diffusion of sulfur vacancy could be observed at a room temperature of 300 K and the lower temperatures because of its higher diffusion energy barrier of over 1 eV and the short ab initio MD simulation time on the order of picoseconds, restricted by the present computational ability. And at the further high 800 K, sulfur vacancy is indeed found to be able to freely diffuse over several steps of lattice constants in the 15-ps MD simulations. It is not found that other SVs could jump over to its nearest neighbor site or form a transition state from 300 K to 800 K, which is quite agreement with the previous conclusion on the small diffusion coefficient.

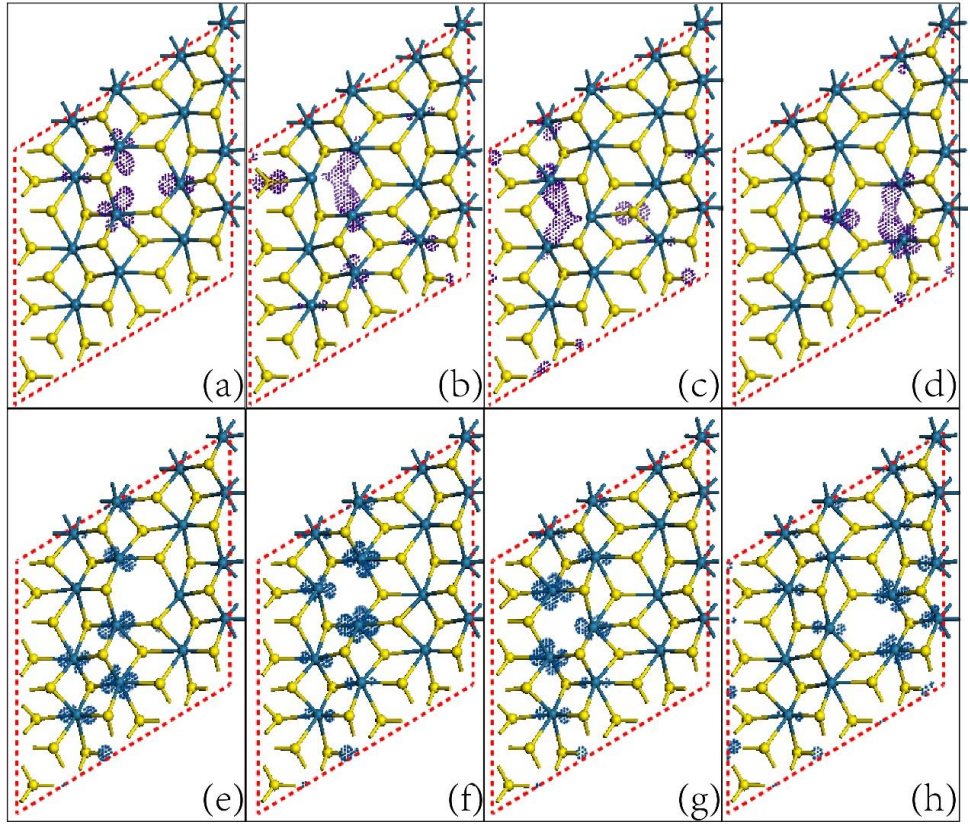

**Figure 4.** (Color on-line) The electron density isosurface of the SVs, the defect state above the Fermi energy: (a): vacancy S1; (b): vacancy S2; (c): vacancy S3; (d) vacancy S4; The defect state below the Fermi energy: (e): vacancy S1; (f): vacancy S2; (g): vacancy S3; (h) vacancy S4;The isovalue is 0.05.

To visibly display the electron redistribution, the electron density of monolayer ReS$_2$ with different SVs are shown in Fig .4. The states above the Fermi energy are localized at the Re atoms around the SVs. The states caused by the vacancies S2, S3 and S4 make the electron clouds overlap sufficiently and lower the energy, while the states caused by S1 do not overlap with each other, which indicates the final ground energy with S1 may be larger than the other SVs, this corresponds to our calculation in Fig. 2(a) and Fig. 2(b). These defect states within the band gap has been investigated in monolayer like MoS$_2$[28]. In monolayer MoS$_2$, the defect states consist of two parts: 1) a single state, made up of mostly $d_{xz}$ and $d_{yz}$ orbitals of the Mo atoms adjacent to the missing S atoms. 2) two degenerate states, made up of mostly the $d_{z^2}$, $d_{x^2-y^2}$ and $d_{xy}$ orbitals of the Mo atoms adjacent to the missing S atoms. Both of the two kinds of states are over Fermi energy and restructure the band gap. In monolayer ReS$_2$ with SV, the defect states above the Fermi energy consist of only the first type which made up of $d_{xz}$ and $d_{yz}$ orbitals, while the 2$^{nd}$ type of states are under the Fermi energy are made up of Re-d orbitals and have no effect on the band gap [as shown in Fig. 4(e)-4(h)].

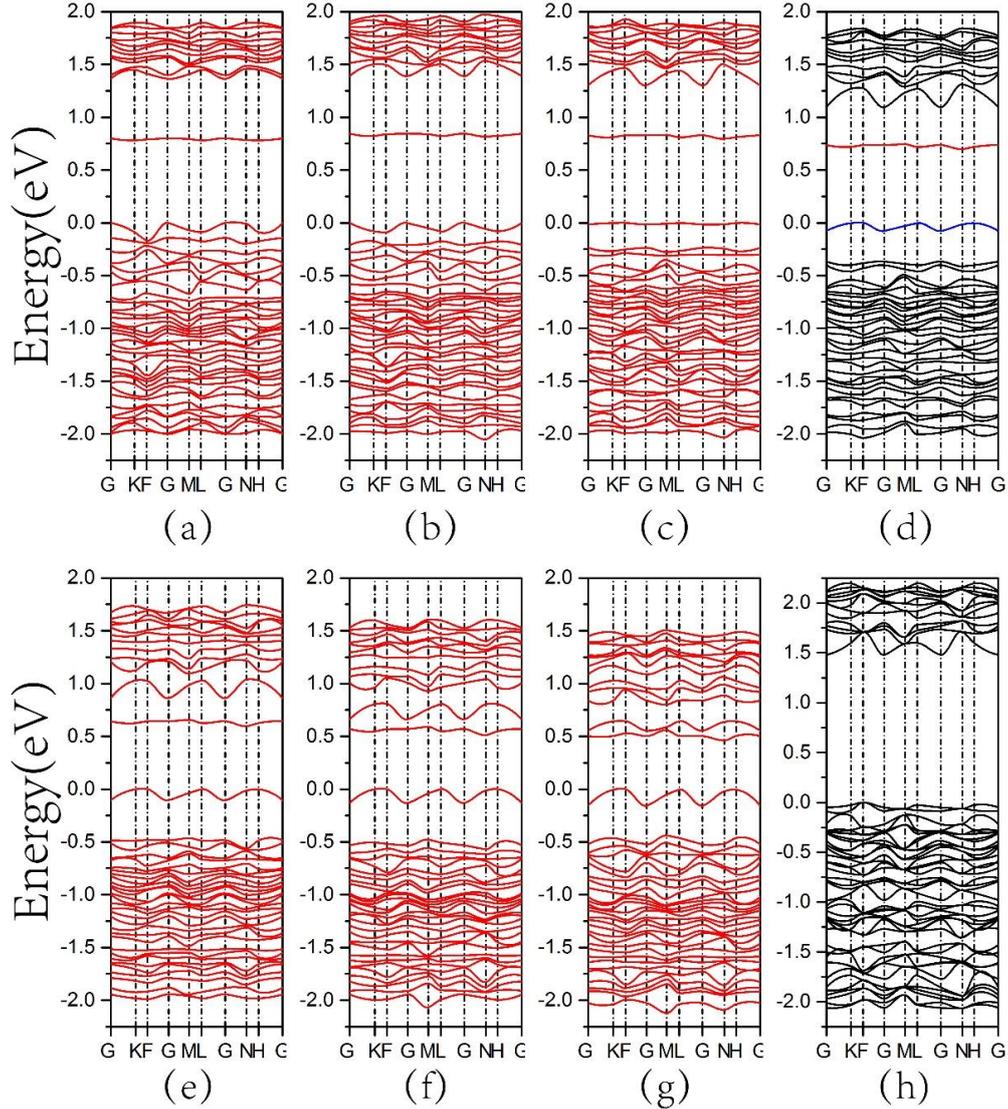

**Figure 5.** (Color on-line) The band structure for the monolayer ReS$_2$ with S4 under different strains along a-direction: (a): -6%, (b): -4 %, (c): -2%, (d): 0%, (e): 2%, (f): 4%, (g): 6%. The two color lines in (d) correspond to two defect states induced by S4 vacancy (Blue:defect state 1; Red: defect state 2). (h): The band structure for the monolayer ReS$_2$ without defect or strain.

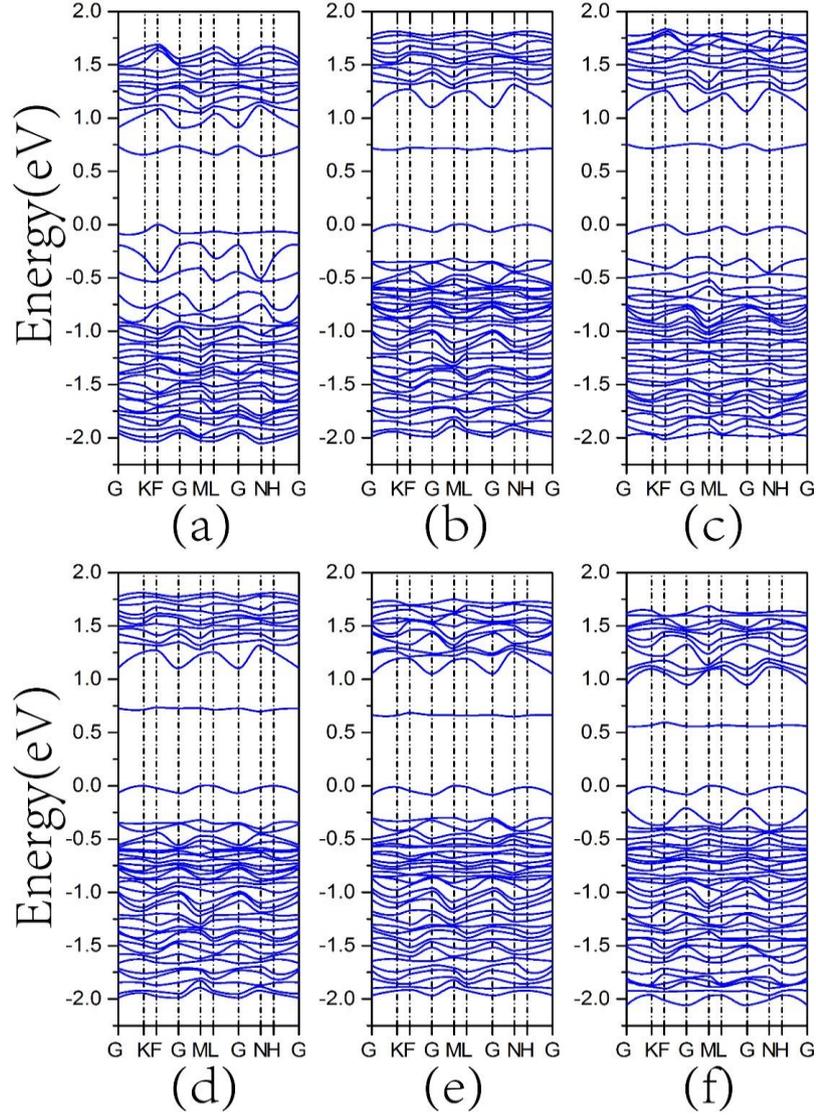

**Figure 6.** (Color on-line) The band structure for the monolayer ReS$_2$ with S4 under different strains along b-direction: (a):-6%; (b):-4%; (c):-2%; (e):2%, (f):4%; (g):6%.

In the previous studies[29], it has been demonstrated that strain plays a key role in tailoring the electronic structure. Formation of different SVs is also discussed in the last section and S4 is the found the most stable SV. The band structure and density of states of S4 under different strains are also investigated. The band structure for monolayer ReS$_2$ without defect is shown in Fig .5(h). The result indicates that the ReS$_2$ monolayer without defect has a direct band gap with 1.53 eV at G-point [0.00, 0.00, 0.00][26]. The band gaps of the defected monolayer ReS$_2$ are largely changed compared with the pristine one. Band structures for monolayer with S4 under different strains are shown in Fig .5 and Fig. 6. The results indicate that the defected ReS$_2$ monolayer has a direct band gap at L-point [0.666, 0.333, 0.000] instead of G-point when there is no external strain. The a and b-direction strains induce indirect band gap with the minimum of conduction band(CBM) shifting from L-point to N [0.000,-0.500, 0.000] in the Brillouin zone (BZ). For the valance band maximum (VBM), when the strains are applied along a-direction, the VBMs do not shift and stay at L-point [0.666,0.333,0.000] in the Brillouin zone at -6%, -4%, -2% and 2%, while the VBMs shift to G-point at 4% and 6%; when the strains are applied along b-direction, the VBMs shift to M-point [0.500,0.000,0.000] at -6%, -4%, -2%, stay at L-point at 2% and 4% and shift to F-point [0.500,0.500,0.000] at 6%. This indicates the band structures are strain dependent and highly anisotropic. For comparison, the strain and direction dependent band gaps of monolayer ReS$_2$ with S4 are shown in Fig. 7(a). For the unstrained monolayer ReS$_2$ with S4, the band gap is 0.697eV. It should be noted that the strains along different directions have different effects on the band gap. As shown in Fig .7(a), it is evident that the band gap of monolayer ReS$_2$ with S4 reach the maximum at -4% and shrink under other strains when the structure is stretched along a-direction. For the strains applied in b-direction, the band gap's maximum is at about 2%.

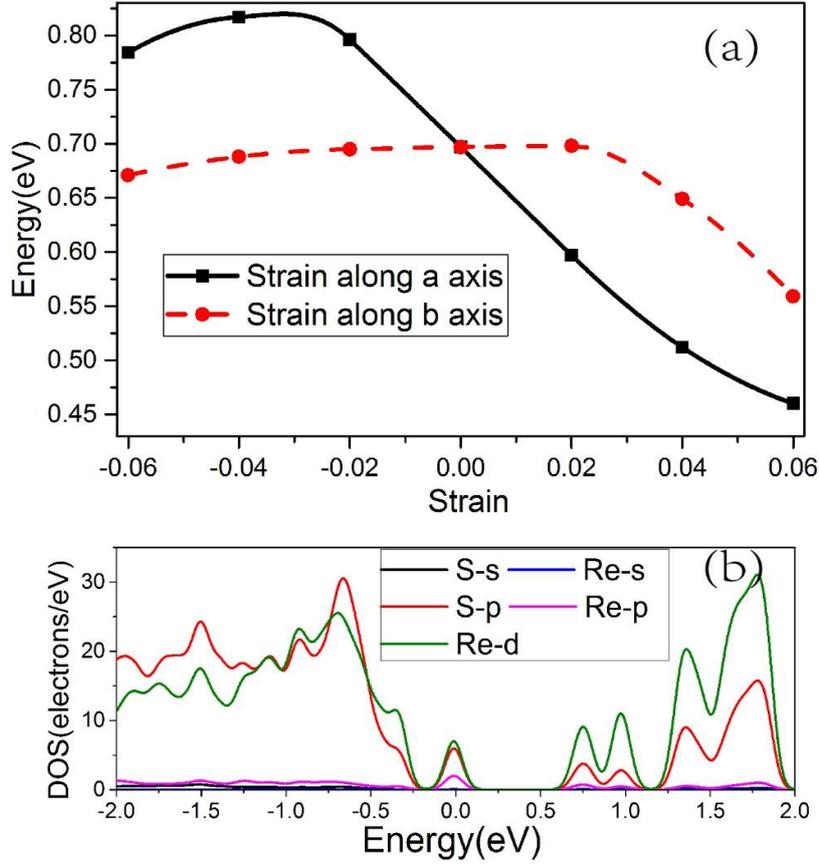

**Figure 7.** (Color on-line) (a) Band Gap of monolayer with S4 vacancy under different strains. (b) Partial density of states (PDOS) of monolayer ReS$_2$ with S4 without strain.

Besides the changes in total energy and band gap, strains also induce change in the partial density of states (PDOS), which is also related to the carrier effective mass ($m^*$): $m^* = \dfrac{\hbar^2}{(\dfrac{\partial^2 E}{\partial k^2})}$, where $\hbar$ is the reduced Planck's constant, E is the energy, and k is the momentum. For the partial density of states of the monolayer with defect, the results are shown in Fig .7(b), Fig. 8 and Fig .9. From Fig .7(b), it is found that the defect states at 0 eV and 0.6 eV consist of hybridization of Re-p states, S-p states and Re-d states. The conduction bands are essentially dominated by Re-d, this echoes with previous conclusions in Fig. 4. The PDOSs under different strains along a and b-direction are shown in Fig. 8 and Fig. 9, it can be found that the S-p and Re-d orbitals are highly splitted at some circumstances (2% along a-direction and -6% along b-direction), indicating the electron delocalization and conductivity enhancement. Also, it is noticed that the highly splitted states in conduction band are reduced when the strain becomes larger along b-direction, which suggests that the localization of electrons[30].

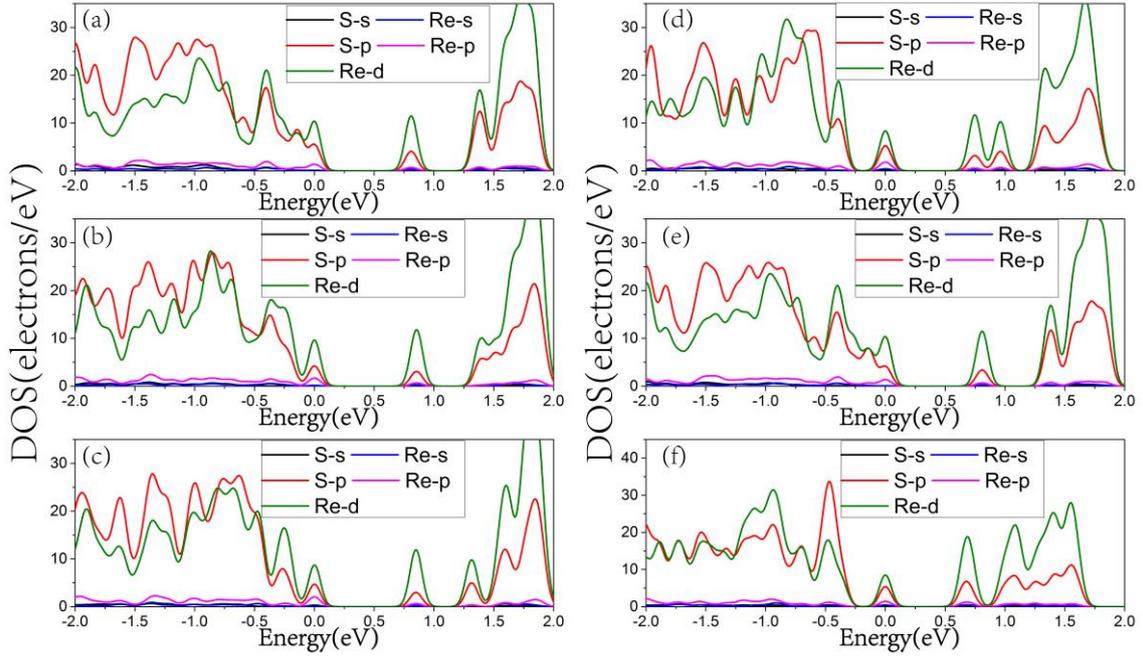

**Figure 8.** (Color on-line) Partial density of states (PDOS) of monolayer ReS$_2$ with S4 with different strains along a-direction. (a) to (f):-6%, -4%, -2%, 2%, 4% and 6%.

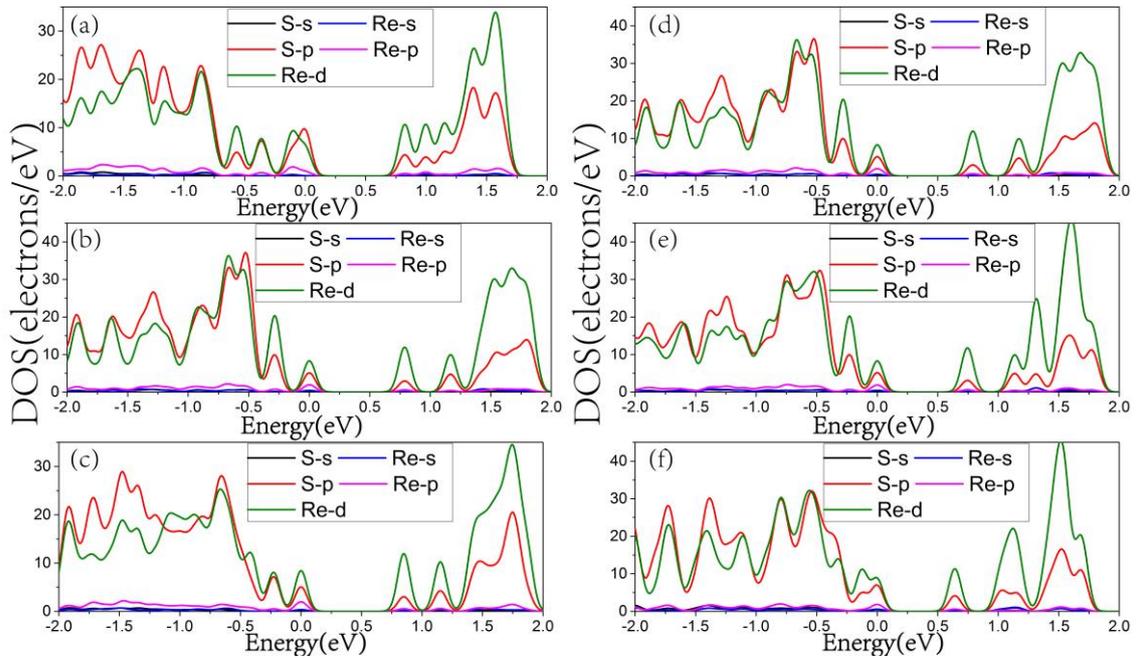

**Figure 9.** (Color on-line) Partial density of states(PDOS) of monolayer ReS$_2$ with S4 with different strains along a-direction. (a) to (f): -6%, -4%, -2%, 2%, 4% and 6%.

In addition, the decreased band gaps can cut down the Schottky barrier height at the ReS$_2$/metal interface in monolayer ReS$_2$ FET devices, which provide potential application in electronic design. The band gaps of monolayer ReS$_2$ with S4s are decrease to about 0.7eV, which are smaller than that of defect-free monolayer ReS$_2$. One reason is that the bonding and anti-bonding orbits are broken when the Sulfur atom is removed. Usually, the energies of broken bonding are increased, whereas the energies of broken anti-bonding are decreased, which brings about the reduction of the band gap. Another reason is that the relaxation of the adjacent surrounding atoms of the vacancy leads to the variation of local DOS, which results in the reduction of the band gap. Additionally, the vacancy charge is unevenly distributed among all neighboring atoms (usually Re atoms) of the vacancy, which also causes the reduction of the band gap.

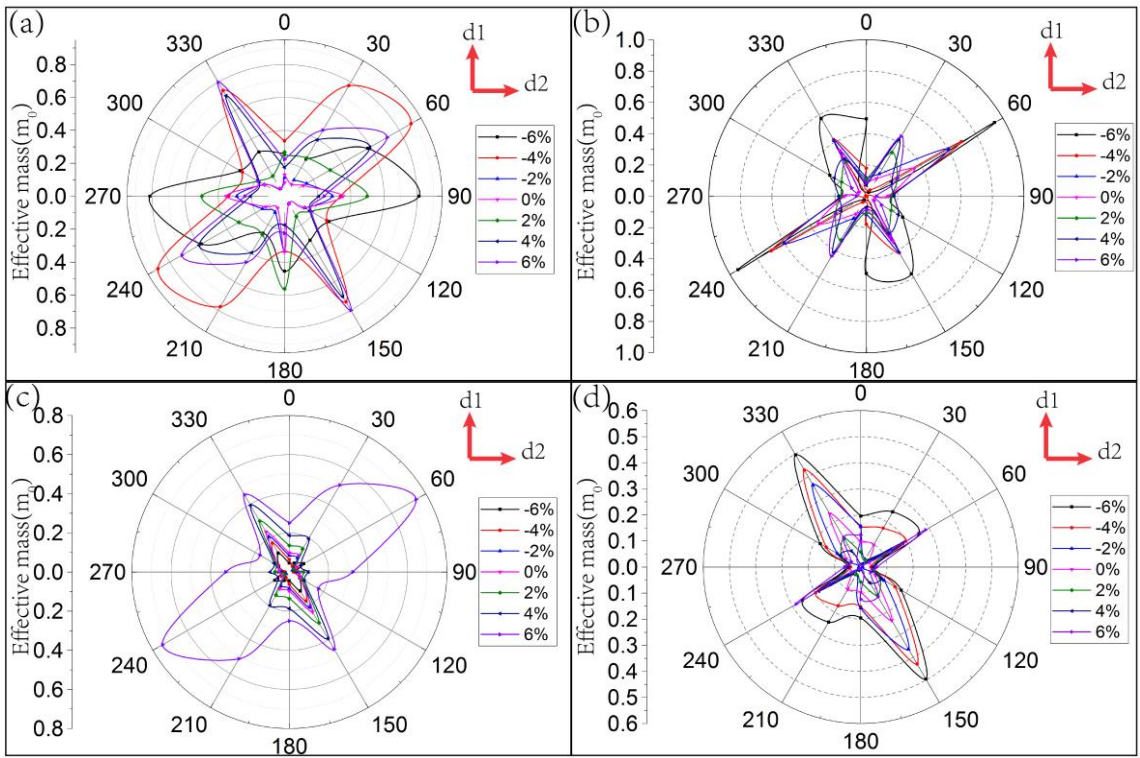

**Figure 10.** (Color on-line) (a) Effective mass of hole (m$_h$) (defect state 1) of monolayer ReS$_2$ with S4 under strain along a direction. (b) Effective mass of hole (m$_h$) (defect state 1) of monolayer ReS$_2$ with S4 under strain along b direction. (c) Effective mass of electron ($m_e$) (defect state 2) of monolayer ReS$_2$ with S4 under strain along a direction. (d) Effective mass of electron ($m_e$) (defect state 2) of monolayer ReS$_2$ with S4 under strain along b direction.

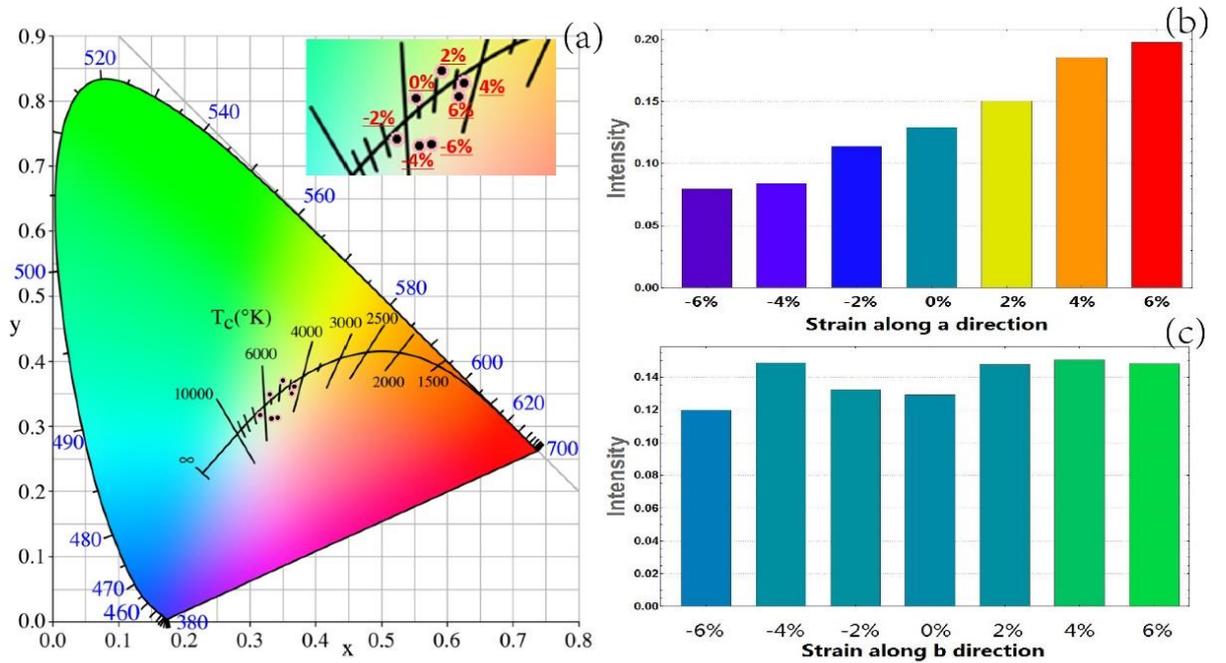

**Figure 11.** (Color on-line) (a) The total refection color under different strains along a-direction. (b) The total peak's color and intensity under different strains along a-direction. (c) The total peak's color and intensity under different strains along b-direction.

Except for energy band gap, the carrier effective mass[31] is another important factor determining the transitions of electrons. Effective masses of charge carriers in materials can be calculated through their energy band structures. Here, the effective masses of carriers versus small strain (−6 %≤ε≤ 6%) along 6 different directions are plotted in Fig. 10. It is found that the response of effective mass to strains is sensitive and anisotropic due to the characteristic puckered structure of ReS$_2$ monolayer with S4. As mentioned above, the strains lead to indirect band gap and shifting of VBM in the Brillouin zone obviously. This direct to indirect band gap transition makes effective mass change significantly when the strains are applied. In Fig. 10, the effective masses of electron ($m_e$) with different strain and angle are shown in Fig. 10(b) and 10(d). The effective masses $m_e$ is changed significantly with applied strain from -6% to 6%. Also, when different direction strains are applied, the monolayer with sulfur defect shows different anisotropic effect: when the a-direction strain is applied, the $m_e$ is relative large at 60° and 150°; for the effective masses of hole ($m_h$), as shown in Fig. 10(a) and 10(c), the a-direction strain induces a higher $m_h$ at 150° comparing to the other direction. It is found that the effective masses at 60° and 150° are more sensitive to the outer strain: the $m_h$ at 60° is 0.88 $m_0$ and 0.13 $m_0$ when -4% and 0% a-direction strains are applied respectively, where $m_0$ is the mass of free electron; the $m_e$ at 150° is 0.43 $m_0$ and 0.07 $m_0$ when -6% and 0% b-direction strains are applied respectively. This large variation indicates the ReS$_2$ monolayer with sulfur defect can be used in strain sensor by engineering carrier transport along the 60° or 150°.

The strains along different directions also change the absorption and reflection spectrum of the monolayer ReS$_2$ with S4. For the reflection spectrum, we investigate the peak of reflection spectrum and the color of the monolayer with defect under different strains. The data of reflected spectrum is shown in Table I and II. The strains along a-direction have more apparent effects on the color of the monolayer. The peak of the reflection spectrum shifts from 407 nm to 660 nm when the strain along a-direction is -6 % and 6 %, while under the strains along b-direction, the peak shifts from 484 nm to 515 nm from -6 % to 6 %. Meanwhile, it is found that strains along a-direction can dramatically increase or decrease the intensity of the peak wavelength. Therefore, the color of the monolayer can be tuned by choosing proper strains. The total refection color is shown in Fig. 11(a), the color's peak and intensity is shown in Fig. 11(b) and 11(c). The larger strain along a-direction induces the increase of the peak's intensity and wavelength. Also, the color temperature of the reflected light shifts from 6000 K at -6% to about 4000 K at 6%.

We also investigate the absorption spectrum of the monolayer, as shown in Fig. 12(a) and Fig. 12(b). The first absorption peak shifts from 73 nm at -6% to 77 nm at 6% when the strains are applied along a-direction. The b- direction's condition is similar. Also, the absorption's intensity [which is the vertical axis in Fig. 12(a) and (b)] shows a symmetrical trend with respect to the $\alpha = 0\%$ (which means no strain is applied on the monolayer ReS$_2$ with S4). The intensity is about 330 cm$^{-1}$ at -6% along a-direction, then increases to about 442 cm$^{-1}$ at 0 % and decreases to 321 cm$^{-1}$ at 6%. Since the magnitude of the absorption is in accordance with the transition probability of the electron, our results suggest that the transition probability of electron at certain frequency is depressed when there are strain[32]. Moreover, the raise of the first absorption peak's wavelength with strain at both a-direction and b-direction suggests that we can know the magnitude of strain applied by detecting the smallest absorption peak.

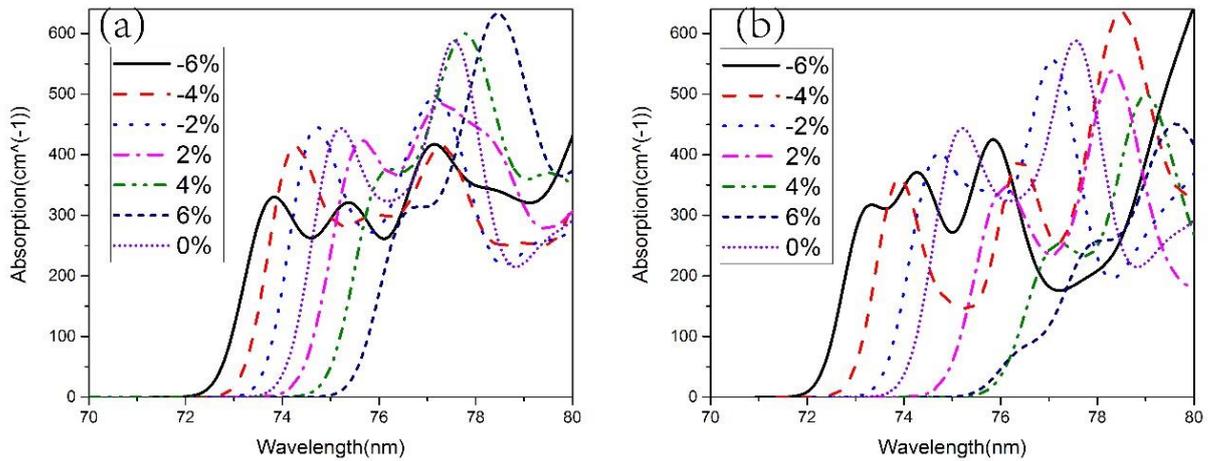

**Figure 12.** (Color on-line) (a) The absorption spectrum under different strain along a-direction. (b) The absorption spectrum under different strain along b-direction.

**Table I.** the coordinate of the color on CIE chromaticity diagram and the peak of monolayer with defect S4 under strain along a-direction.

| Strain | CIE_X | CIE_Y | Peak(nm) | Peak Intensity |
| --- | --- | --- | --- | --- |
| -0.06 | 0.3433 | 0.3128 | 407 | 0.07973 |
| -0.04 | 0.3329 | 0.3117 | 443 | 0.08396 |
| -0.02 | 0.3156 | 0.3175 | 458 | 0.11348 |
| 0 | 0.3307 | 0.3489 | 488 | 0.12912 |
| 0.02 | 0.351 | 0.3706 | 590 | 0.15043 |
| 0.04 | 0.3685 | 0.3616 | 615 | 0.18514 |
| 0.06 | 0.3636 | 0.3503 | 660 | 0.19741 |

**Table II.** The coordinate of the color on CIE chromaticity diagram and the peak of monolayer with defect S4 under strain along b-direction.

| Strain | CIE_X | CIE_Y | Peak(nm) | Peak Intensity |
|---|---|---|---|---|
| -0.06 | 0.3221 | 0.3525 | 484 | 0.12007 |
| -0.04 | 0.3392 | 0.3662 | 490 | 0.14865 |
| -0.02 | 0.3298 | 0.3473 | 491 | 0.13207 |
| 0 | 0.3307 | 0.3489 | 488 | 0.12912 |
| 0.02 | 0.3385 | 0.3657 | 490 | 0.14769 |
| 0.04 | 0.3486 | 0.3939 | 508 | 0.15054 |
| 0.06 | 0.3426 | 0.401 | 515 | 0.14822 |

## **CONCLUSION**

In summary, we have explored the strain effect on the properties of the defected ReS$_2$ monolayer by using first-principle calculations. The monolayer ReS$_2$ with S4 is the most stable comparing to other structures and the high diffusion barrier prevent other structure from migrating to it. The ReS$_2$ monolayer exhibits different electronic properties along different directions: a-direction and b-direction strains induce indirect band gap and change the band gap. The calculation of partial density of states implies that the defect states are mainly originated form Re-p, Re-d and S-p orbitals. Also, the effective mass can be significantly improved and shift from negative to positive by asymmetric deformation induced by strains. The reflection spectrum is more sensitive when the strains are applied along a-direction than b-direction. Those results indicate that defected monolayer ReS$_2$ has promising applications in nanoscale strain sensor and conductance-switch FETs.

AUTHOR INFORMATION

Corresponding Author

*lzliu@nju.edu.cn (LZ Liu)

Notes

The authors declare no competing financial interest.

ACKNOWLEDGMENTS

This work was supported by National Natural Science Foundation of China (Nos. 11404162). We also acknowledge the computational resources provide by High Performance Computing Centre of Nanjing University